%% file: main.tex
\documentclass[aps,prl,amsmath,amssymb,reprint,superscriptaddress]{revtex4-2}
\usepackage[T1]{fontenc}
\usepackage{graphicx}
\usepackage{natbib}

\usepackage[capitalize]{cleveref}
\input{commands.tex}

\begin{document}

\title{Shape-Driven Caging Dynamics of Hard Polygons}
\author{Vyas~Ramasubramani}
\email{vramasub@umich.edu}
\affiliation{Department of Chemical Engineering, University of Michigan, Ann Arbor, MI 48109}

\author{Thi~Vo}
\email{thiv@umich.edu}
\affiliation{Department of Chemical Engineering, University of Michigan, Ann Arbor, MI 48109}

\author{Joshua~A.~Anderson}
\email{joaander@umich.edu}
\affiliation{Department of Chemical Engineering, University of Michigan, Ann Arbor, MI 48109}

\author{Sharon~C.~Glotzer}
\email[Corresponding Author: ]{sglotzer@umich.edu}
\affiliation{Department of Chemical Engineering, University of Michigan, Ann Arbor, MI 48109}
\affiliation{Department of Physics, University of Michigan, Ann Arbor, MI 48109}
\affiliation{Biointerfaces Institute, University of Michigan, Ann Arbor, MI 48109}

\begin{abstract}
    The fundamentals of Brownian motion have been largely well understood since the early 20th century, with most recent additions focusing on understanding anomalous diffusion via rescaling of drag coefficents. That focus emphasizes long-time dynamic behavior, but recent results indicate that additional, secondary modes are also present at short and intermediate times in fluids of anisotropic particles.
    Here, we study the dynamics of a representative family of nearly-hard \ngons.
    Using \acl{md} simulations, we study a distinct form of caging only present in anisotropic systems.
    We show that this caging behavior emerges in the \acl{msd} of \ngons\ at intermediate particle volume fractions. We then develop an extended Langevin theory directly coupling translational and rotational motion as a function of the relative anisotropy for different \ngons\ that predicts the observed caging behavior. Extending our theory to incorporate secondary, off-phase cross-correlations between particles further enables the prediction of both translational and rotational relaxation times of the system.
\end{abstract}

\maketitle

The seminal contributions of Einstein and Smoluchowski in the early 1900s mark the beginning of the modern study of Brownian motion. Their works established a relationship between the long-time diffusion constant of a particle and its friction coefficient -- commonly referred to as the \ac{se} relation -- and provided a means to estimate the diffusion coefficient in a fluid via observation of a particle's \ac{msd}, a crucial step in validating the atomic theory of matter \cite{Einstein1905,VonSmoluchowski1906}. Their findings opened the floodgates to follow-up works such as the Langevin equation \cite{Langevin1908} and the incorporation of Brownian motion into formal theories of stochastic processes \cite{Uhlenbeck1930} that have since made the topic ubiquitous in mathematics and the sciences. 

Extensions were made to generalize the theory of translational Brownian motion to rotational diffusion via rescaling the diffusion constant \cite{Einstein1906} as well as direct coupling between rotational and translational diffusion for ellipsoidal particles \cite{Perrin1934,Perrin1936}, culminating in the widely used \ac{sed} relation \cite{Mazza2007}. Further generalizations were proposed for anisotropic particles \cite{Brenner1965,Brenner1967}, but these results remained relatively obscure outside of a few theoretical studies \cite{Wegener1981,Harvey1980,Dickinson1985,Kholodenko1995} until they were experimentally validated using high resolution digital video microscopy \cite{Han2006}. A common assumption employed throughout the above works is the idea that long-time diffusion of aspherical particles is identical to that of spherical particles and dynamical variations arising from particle anisotropy only manifest in the short-time regime. As such, previous theoretical developments have particularly focused on the idea of rescaling the long-time diffusion constant of an anisotropic system, leaving the core differential form of the Langevin equation unaltered \cite{Brenner1965,Brenner1967,Wegener1981,Harvey1980,Dickinson1985,Kholodenko1995}.

Recent findings using modern computational and high resolution experimental imaging techniques, however, have provided evidence for the emergence of intermediate caging regimes in the measured \acp{msd} of anisotropic particles \cite{Taloni2015,Mazza2007,Chakrabarty2013,Chakrabarty2014a,Chakrabarty2016,Wang2018,Hou2019,Li2019}. Due to the presence of particle anisotropy, such regimes are associated with particle shape and distinct from traditional caging observed in crowded systems of spheres. These findings point at interesting dynamics manifesting from the interplay between instantaneous rotational and translational motions of anisotropic particles that, upon dissection, could reveal new physics. Doing so, however, necessitates a reconsideration of the form of the classical Langevin equation and subsequent relaxation of the assumption that rescaling the long-time diffusion constant for anisotropic systems is sufficient for capturing the relevant dynamical behaviors. This is because solutions of the classical Langevin equation can only predict a short-time ballistic and long-time diffusive behavior. As such, any intermediate caging modes cannot be captured using the approximations inherent in its derivation.

Previous efforts aimed at understanding the instantaneous behaviors of dynamical systems culminated in well-established and validated theories of time correlation functions as well as mode coupling analyses that are now widely employed in the study of liquids \cite{Hansen2006,Gotze2009,Reichman2005,Zwanzig2001}. Such theories take advantage of the projection-operator formalism and memory functions for predicting dynamical responses across a wide range of different systems. Examples include standard Lennard-Jones \cite{Levesque1970,Gaskell1978}, charged \cite{Hansen1975}, molecular \cite{Bien1981,Pollock1981}, and even polymeric  \cite{Mirigian2015,Xie2020} liquids. However, due to the complexity arising from interactions intrinsic to real liquids, these approaches often employ input from computer simulations to define the decay of the relevant memory functions or self-consistent techniques in cases where phenomenological simplifications can be made to approximate their functional forms \cite{Levesque1970,Gaskell1978,Hansen1975,Bien1981,Pollock1981,Mirigian2015,Xie2020}. This is particularly true for systems of anisotropic particles where analytical treatments opt to separate translational and rotational behaviors and resort to simulation inputs to study the effects of coupling \cite{Misra2017,Lang2013,Charbonneau2014}. As such, simple solutions akin to those produced by the Langevin equation remain elusive for anisotropic systems.

In this work, we study a model system of hard anisotropic particles such that all deviations away from classical Brownian motion can be attributed to particle shape and the effects of anisotropy on collisions. In this regard, this system serves as an ideal testing ground for developing the extensions to theories of Brownian motion. Previous studies have shown that hard anisotropic particles develop statistical, effective \acp{def} upon crowding to maximize the system entropy \cite{VanAnders2014,VanAnders2014a}. Our results indicate that these \acp{def} lead to the emergence of a caging mode at intermediate particle volume fraction that arises solely due to particle shape, distinct from other modes of caging for systems of isotropic particles. We propose a generalized Langevin theory that stems from simplifications of the previously mentioned projection-operator approaches in the limit of hard, otherwise non-interacting, anisotropic particles and show that the new theory captures the volume fraction dependent emergence of this shape-driven caging behavior. We further show that an extension of this theory predicts the translational and rotational relaxation times for different shapes, including unexpected non-monotonic behavior with respect to the number of sides of the polygon that hints at the presence of more complex, higher-order effects.

\paragraph{MSD of Hard Polygons} ---
We simulated nearly-hard polygons using a recently developed anisotropic analogue to the \ac{wca} potential \cite{Chandler1983a,Ramasubramani2020b}. Molecular dynamics simulations were conducted in the \textit{NVT} ensemble with a Nos\'{e}-Hoover thermostat \cite{Martyna1994} using the \hoomd\ simulation package \cite{Anderson2020}. We consider unit area regular \ngons\ where $n$ ranges from $3-9$ across a wide range of particle volume fractions $\phi$ corresponding to fluid phases for all polygons. Each simulation contains $N=1024$ particles run using \hoomd's reduced units of $\epsilon = 1$, temperature $T^{*} = 1$, and a timestep of $10^{-4}$.

\begin{figure}[t]
    \includegraphics[width=1.05\columnwidth]{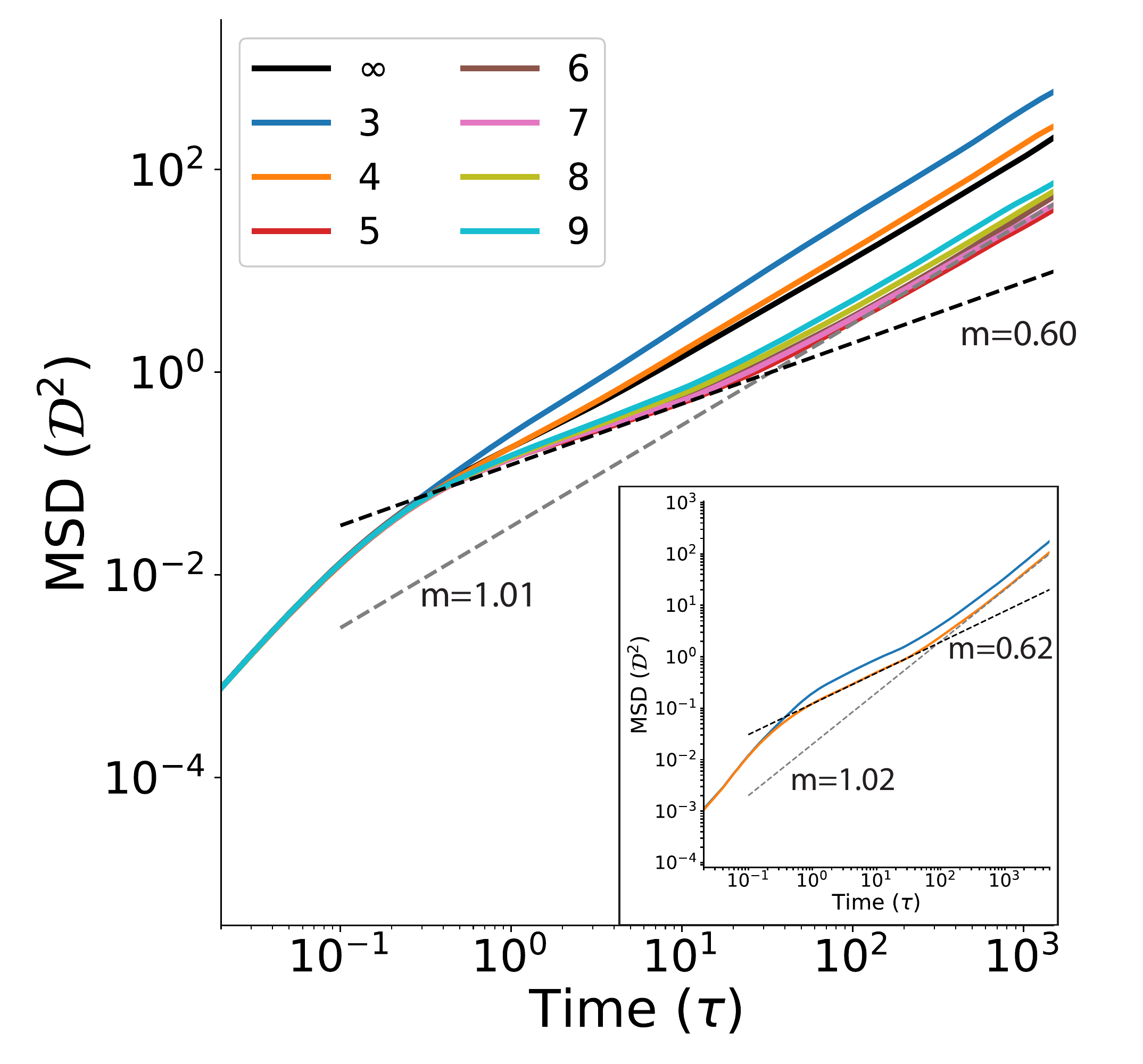}
    \caption{
        The \ac{msd} of polygons at $\phi = 0.68$.
        At this intermediate particle volume fraction the \ac{msd} exhibits a caging regime (dashed black line, slope $m=1.01$) prior to the onset of diffusion (dashed gray line, slope $m=0.60$) for polygons.
        The solid black line is the \ac{msd} of disks.
        Although this caging regime is present for all polygons, it appears at different particle volume fractions for triangles and squares (inset) than for other polygons.
        The plot is shown in the reduced distance and time units used by \hoomd.
        The \ac{msd} was computed using the \freud\ analysis package \cite{Ramasubramani2020}.
    }
    \label{fig:IntermediateMSD}
\end{figure}

While the \acp{msd} of these shapes all collapse onto the classical limit for dilute fluids (Section I, SI), the \acp{msd} at intermediate particle volume fractions clearly show a plateau indicating caging behavior that depends on the number of sides of the polygon, appearing around $\phi=0.68$ for $n \geq 5$ and $\phi=0.75$ for triangles and squares (see \cref{fig:IntermediateMSD}). Although hard disk fluids also exhibit various forms of caging, the particle volume fraction at which we observe this caging transition is evidently a function of particle shape, suggesting that this particular caging behavior is influenced by rotation-translation coupling, a general feature of anisotropic systems. The projection operator formalism \cite{Zwanzig2001} can be used to develop a generalized Langevin equation that accounts for this coupling, yielding an equation we can solve for an \ac{msd} that directly embeds rotation-translation coupling.
We start with the standard equation of motion of observables \vb{A}:

\begin{equation}
    \frac{\partial}{\partial t} \mathbf{A}(t) - i \mathbf{\Omega} \cdot \mathbf{A} + \int_{0}^{t} \mathbf{M}(t - s) \cdot \mathbf{A}(s) ds = \mathbf{f}(t)
    \label{eq:motion}
\end{equation}

\noindent where $\mathbf{\Omega}$ is the frequency matrix, $\mathbf{M}$ is the memory kernel, and $\mathbf{f}$ is the random noise \cite{Zwanzig2001}. The classical Langevin equation results from focusing on the linear velocity as the observable of interest. However, anisotropic particles require a multivariate observable of position ($\vb{r}$), velocity  ($\vb{v}$), and angular velocity  ($\vb{l}$), such that $\vb{A} = \left[ \vb{r}, \vb{v}, \vb{l}_p \right]$, where $\vb{l}_p$ is the projected form of $\vb{l}$, ensuring orthogonality. Applying the projection framework to this set of observables gives a system of three, coupled differential equations (Section II, SI). To parallel the approach employed in the derivation of the classical Langevin equation, we focus only on the velocity equation:

\begin{equation}
	m \dot{\mathbf{v}}(t) + \xi \mathbf{v}(t) + \gamma_c \left( \alpha / I_n \right)^{-1/2} \mathbf{v}^2(t) = f(t)
	\label{eq:differential}
\end{equation}

\noindent where $I_n$ is the moment of inertia of any given regular $n$-gon, $f$ is the fluctuating force, $\alpha$ is a proportionality constant relating linear and angular velocities, $\xi$ is a friction coefficient, and $\gamma_c$ is a translational-to-rotational coupling constant that encodes a relationship between translational and rotational friction.

In deriving this expression we have focused on the real components (i.e. $i \mathbf{\Omega} \cdot \mathbf{A} = 0$) and assumed a constant memory kernel. The same approximations were used for the classical Langevin equation; therefore, \cref{eq:differential} can be interpreted as an anisotropic generalization of the classical Langevin equation. In addition, to eliminate terms in angular velocity, we derived a relationship between the linear and rotational momenta of polygons (Section II, SI), which results in the unfamiliar term in $\mathbf{v}^2$. In the limit of $\gamma_c \rightarrow 0$ (i.e. no rotation-translation coupling), we recover the classical Langevin equation. Like in the classical Langevin equation, most of these terms can be evaluated analytically within the framework of this theory (see SI). While individual constants can be analytically computed, no exact solution exists for \cref{eq:differential}; however, an approximation can be obtained via Taylor expansion about the intermediate time regime to give the following functional form for the \ac{msd} of polygons (Section III, SI):

\onecolumngrid

\noindent\rule{16cm}{0.4pt}

\vspace{-4.5mm}

\begin{equation}
	\begin{split}
		\langle r^2 (t) \rangle = \frac{1}{4\gamma^2} \Bigg\{ B_1 + B_2 e^{-\xi t} \Bigg[ e^{-\xi t} + \left( \frac{1}{\xi^2 e^2} \right) t + \frac{2}{\xi^2} \Bigg] \Bigg\}^2 + \left( \frac{B_3}{\xi} \right) t + \frac{1}{2} \frac{B_3^2 \gamma^2}{\xi^4} \hspace{9mm} \\
		B_1 = 4 \left[ \frac{B_3^2 \gamma^4}{\xi^4} \right]^2 + B_2 \left[ 1 + \frac{2}{\gamma^2} \right], \hspace{2mm} B_2 = - \frac{1}{2} \left[ \frac{\xi^7}{B_3^3 \gamma^6} \right] \left[ 2 \xi + \frac{2}{\xi} - \frac{1}{\xi^2 e^2} \right]^{-1} \hspace{-4mm} , \hspace{3mm} B_3 \sim \frac{\langle m v^2 (t) \rangle}{m}
	\end{split}
	\label{eq:msd_solution}
\end{equation}

\begin{flushright}
	\noindent\rule{16cm}{0.4pt}
\end{flushright}

\twocolumngrid

\noindent where we have defined $\gamma = \gamma_c \left( \alpha / I_n \right)^{-1/2}$ and $ \langle r^2 (t) \rangle \equiv \langle \left( r(t) - r(t_o) \right)^2  \rangle$. \Cref{eq:msd_solution} contains the usual linear term at long times, but expanding the exponential yields a secondary linear regime; combining both gives

\begin{equation}
	\langle r^2 (t) \rangle \sim \Bigg\{ \frac{B_3}{\xi } + \frac{\left( e^{-2} - 2 \xi \right) \left( 2 \xi^{-1} - B_1 B_2^{-1} \right)}{2 B_2^2 \gamma^2 \xi^2} \Bigg\} t
	\label{eq:linearmsd}
\end{equation}

\begin{figure}[b!]
    \centering
    \includegraphics[width=\columnwidth]{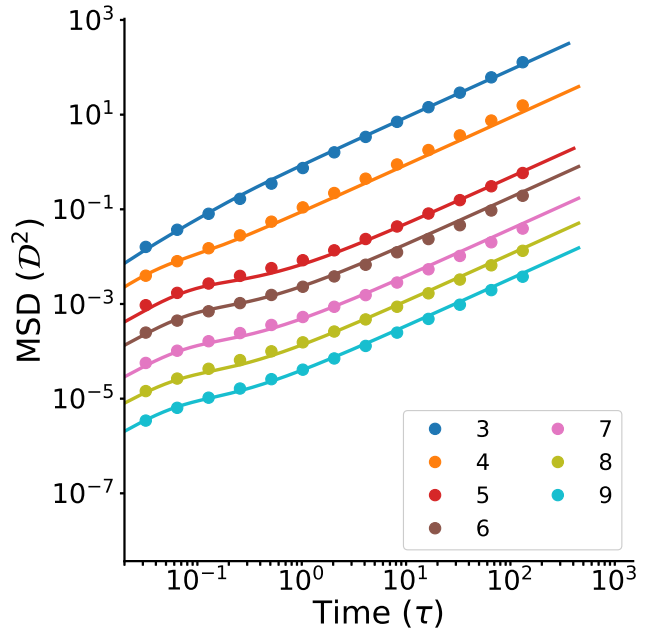}
    \caption{
        Comparison of measured MSD from \cref{fig:IntermediateMSD} with theory prediction at $\phi = 0.68$; all systems are in a fluid phase.
        Data points are from simulation, lines indicate theory predictions.
        \acp{msd} are artificially offset to clearly show comparison between theory and simulation.
    }
    \label{fig:IntermediateMSDWithTheory}

\end{figure}

The first term on the right hand side of \cref{eq:linearmsd} is analogous to SE and the second term results from coupling between rotation and translation. Therefore, the long-time behavior of \cref{eq:msd_solution} is interpreted as perturbations about the SE limit due to intermediate rotational-translational coupling, producing a net ``average'' diffusion constant. Phenomenologically, the long-time limit converges to the commonly employed approach of rescaling the long-time diffusion constant. Plugging $B_1$ and $B_2$ into \cref{eq:linearmsd}, the solution also converge to the \ac{se} limit for $\gamma \rightarrow 0$. At short times, the exponential terms vanish and the classical $t^2$ ballistic solution is recovered. At intermediate times, however, the terms involving decaying exponentials become non-negligible, providing the source for the observed caging regime. As shown in \cref{fig:IntermediateMSDWithTheory}, our theory predicts the shape-driven caging behavior observed in the simulations across different shapes, including its dependence on particle volume fraction (Fig. III.1, SI).

\begin{figure}[t!]
	\centering
	\includegraphics[width=\columnwidth]{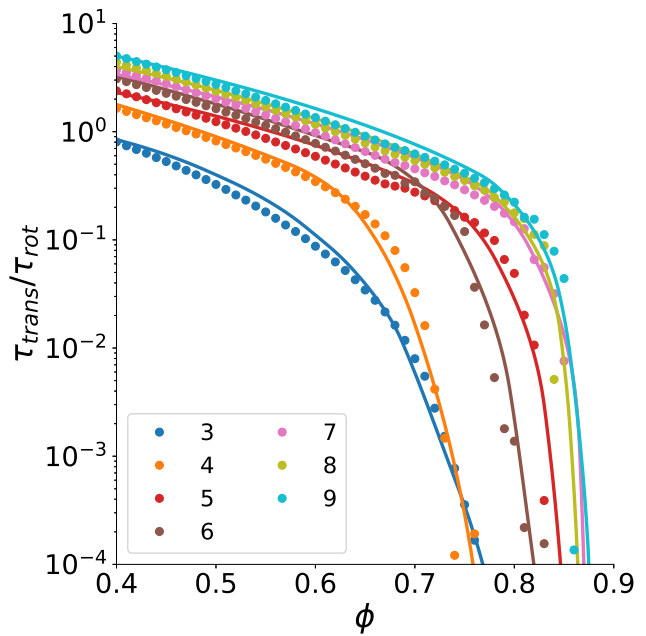}
	\caption{
		Ratio of translational to rotational relaxation time as a function of particle volume fraction for different $n$-gons.
		$\tau_{trans} \tau_{rot}^{-1}$ as predicted from \cref{eq:tau_trans,eq:tau_rot} (solid lines) overlaid with simulation (data points).
	}
	\label{fig:RelaxationRatio}
\end{figure}

\paragraph{Estimating Relaxation Times} ---
To gain deeper insight into the dynamical behavior of this family of $n$-gons, we extracted the relaxation times for translation \taut\ and rotation \taur\ from simulation (data points in \cref{fig:RelaxationRatio}). Inspection of  the data points in \cref{fig:RelaxationRatio} reveals that \tauratio\ drops off more quickly for $3$-, $4$-, and $6$-gons than for other polygons, with the data for the $6$-gons crossing that of the $5$-gons. One possible explanation lies in the emergence of strong \acp{def} between polygons upon increased crowding. Previous computational studies have shown that, while  \acp{def} emerge in all crowded systems of polygons, only $3$-, $4$-, and $6$-gons exhibit strong \acp{def} while still in the disordered fluid phase \cite{VanAnders2014,VanAnders2014a,Anderson2017c}. Strong \acp{def} lead to orientational alignment between particles that then drive the observed rapid arrest of rotational relative to translational dynamics. Additionally, the emergence of strong \acp{def} induces effective attractions between particles, resulting in long-lived local motifs, which can be interpreted in terms of caged dynamics \cite{saltzmanschweizer,Mirigian2015}. Our theoretical treatment thus far does not take caging into account, and so not surprisingly is unable to properly explain the data (Fig. III.2, SI). This motivates us to reintroduce the frequency term $i \mathbf{\Omega} \cdot \mathbf{A}$ into the governing equations. This term measures the off-phase cross-correlations between particles and captures local fluctuations arising from caging. Such cross-correlations are of secondary importance for MSD predictions since \cref{eq:msd_solution} is interested in the emergent, bulk effect of rotation-translation coupling, and thus the simpler expression is sufficient to reproduce the data. However, dissecting the interplay between \taut\ and \taur\ seeks to decouple rotational and translational motions, necessitating the reintroduction of relevant frequency terms. Doing so results in the following modification of \cref{eq:differential} 

\small

\begin{equation}
	m \dot{\mathbf{v}}(t) + \left[ \xi + \eta \left( \alpha / I_n \right)^{-1/4}  \right] \mathbf{v}(t) + \gamma_c \left( \alpha / I_n \right)^{-1/2} \mathbf{v}^2(t) = f(t)
	\label{eq:differential2}
\end{equation}

\normalsize

\noindent where $\eta$ is a term encapsulating the projection of a polygon's rotational mode onto its positional mode (Section II, SI). The solution to \cref{eq:differential2} is analogous to \cref{eq:msd_solution} with $\xi$ replaced with $\xi_{\omega} \sim \xi + \eta \left( \alpha / I_n \right)^{-1/4}$. To derive explicit expressions for \taut\ and \taur\, we employ the following scaling arguments. We define \taur\ as the time when the the rotation-translation coupling terms ($B_1$ and $B_2$) are dominant and \taut\ as the time when the standard diffusive term $B_3$ dominates in the solution to \cref{eq:differential2} (Section III, SI). The functional forms for the relaxation times are:

\small

\begin{equation}
	\tau_{trans} = \xi_{\omega}^{-1} \mathcal{W} \left( \frac{B_1 B_2^{-1} - 1 + \left( \xi_{\omega}^{-1} - \xi_{\omega}^{-2} \right) \left( e^{-2} + 2 \right)}{ \gamma B_2^2 B_3 \xi_{\omega}^{-2} } \right)
	\label{eq:tau_trans}
\end{equation}

\begin{equation}
	\tau_{rot} = \xi_{\omega}^{-1} \mathcal{W} \left( \frac{B_1 B_2^{-1} - 1 + \left( \xi_{\omega}^{-1} - \xi_{\omega}^{-2} \right) \left( e^{-2} + 2 \right)}{ \left( e^{-2} - 2 \xi_{\omega} \right) \left( 2 \xi_{\omega}^{-1} - B_1 B_2^{-1} \right) \xi_{\omega}^{-3} }  \right)
	\label{eq:tau_rot}
\end{equation}

\normalsize

\noindent where $\mathcal{W}(x)$ defines the Lambert W-function \cite{Corless1996}.

Computing \tauratio\ using \cref{eq:tau_trans,eq:tau_rot} for all \ngons\ across different particle volume fractions shows good agreement between theory and simulation (\cref{fig:RelaxationRatio}), suggesting that frequency effects can serve as a first-order perturbative correction for relaxation behavior within the secondary regime. Comparisons between theory and simulation for \taur\ and \taut\ individually are shown in Fig. III.3 and III.4 of the SI, respectively. The resulting estimate of \tauratio\ captures the expected limiting behaviors of the system reasonably well: \tauratio\ is very large at low $\phi$ because velocity correlations die off slowly due to the low frequency of collisions. Conversely, \tauratio\ decays to zero at high $\phi$ because rotations in this system become arrested (drastically augmenting $\tau_{rot}$) while velocities are effectively random due to the large number of local collisions. The theory even correctly predicts the crossing of the data for $5$-gons and $6$-gons around $\phi = 0.74$ observed in the simulation data. Without the frequency term, no crossing is predicted, as shown in Fig III.2 in the SI. 

It is worthwhile noting here that there appears to be a consistent slight overestimation of \tauratio\ in the low $\phi$ limit prior to the onset of caging ($\phi \le 0.6$). We suspect that this results from our choice to approximate the memory kernels ($\mathbf{M}$) in \cref{eq:motion} as constants rather than a convolutional integral with their corresponding position, velocity, or angular velocity terms. Doing so underestimates the effects of cross-correlations and results in an augmented $\tau_{trans}$. Addressing these secondary effects involves dissecting the full memory kernel $\mathbf{M}$ which goes beyond our current aim of developing an analytical extension to the classical Langevin equation. 

The observed separation of timescales for simple systems of regular polygons points at the presence of rich, unexplored dynamical behavior within the family of anisotropic shapes. Even more complex behaviors may lie in wait upon extension to 3-dimensional systems of polyhedra. Proper understanding and elucidation of such hidden interactions not only enrich our understanding of anisotropic dynamics, but also provide crucial insights needed for future assembly engineering applications of shaped building blocks.

\vspace{1mm}

This research was supported in part by the National Science Foundation, Division of Materials Research Award \# DMR 1808342 and by the Department of the Navy, Office of Naval Research under ONR award number N00014-18-1-2497. 
V.R. acknowledges the 2019-2020 J. Robert Beyster Computational Innovation Graduate Fellowship from the College of Engineering, University of Michigan.
This research was supported in part through computational resources and services supported by Advanced Research Computing at the University of Michigan, Ann Arbor.
This work used the Extreme Science and Engineering Discovery Environment (XSEDE) \cite{Towns2014a}, which is supported by National Science Foundation grant number ACI-1548562; XSEDE award DMR 140129.
This research used resources of the Oak Ridge Leadership Computing Facility, which is a DOE Office of Science User Facility supported under Contract DE-AC05-00OR22725.
Hardware provided by NVIDIA Corp. is gratefully acknowledged.

\bibliography{main}

\end{document}

%% file: commands.tex
\usepackage[nonumberlist,acronym,xindy, nomain, shortcuts]{glossaries}
\newacronym{msd}{MSD}{mean squared displacement}
\newacronym{gjk}{GJK}{Gilbert-Johnson-Keerthi}
\newacronym{md}{MD}{molecular dynamics}
\newacronym{wca}{WCA}{Weeks-Chandler-Anderson}
\newacronym{mc}{MC}{Monte Carlo}
\newacronym[longplural={potentials of mean force and torque}]{pmft}{PMFT}{potential of mean force and torque}
\newacronym{pmf}{PMF}{potential of mean force}
\newacronym{se}{SE}{Stokes-Einstein}
\newacronym{sed}{SED}{Stokes-Einstein-Debye}
\newacronym{ode}{ODE}{ordinary differential equation}
\newacronym{def}{DEF}{directional entropic force}

\newcommand{\vb}[1]{\ensuremath{\mathbf{#1}}}
\newcommand{\ngon}{$n$-gon}
\newcommand{\ngons}{\ngon s}

\newcommand{\taut}{\ensuremath{\tau_{trans}}}
\newcommand{\taur}{\ensuremath{\tau_{rot}}}
\newcommand{\tauratio}{\ensuremath{\tau_{trans}/\tau_{rot}}}
\newcommand{\hoomd}{HOOMD-blue}
\newcommand{\freud}{\texttt{freud}}

\crefname{figure}{Fig.}{Figs.}

\newcommand{\fig}[1][]{%
    \textbf{Add Figure
    \ifthenelse{\equal{#1}{}}{}{: #1}}%
}

